\documentstyle[twoside,fleqn,espcrc2,psfig]{article}

\newcommand{\AmS}{{\protect\the\textfont2
  A\kern-.1667em\lower.5ex\hbox{M}\kern-.125emS}}

\title{BeppoSAX observations of PKS 0528+134}

\author{G. Ghisellini, G. Tagliaferri, L. Costamante, L. Maraschi; 
        \address{Brera Astronomical Observatory, V. Bianchi, 46 Merate, Italy}
        A. Celotti, G. Fossati; \address{SISSA, Trieste, Italy}
        L. Bassani, M. Cappi F. Frontera, E. Pian; 
        \address{TESRE Bologna, Italy}
        A. Comastri;\address{Osserv. Astron. of Bologna, Italy}
        M. Cavallone, G. De Francesco, L. Lanteri, C.M. Raiteri, G. Sobrito, 
        M. Villata; \address{Osserv. Astron. di Torino, Italy}
        S. Giarrusso, B. Sacco; \address{IFCAI/CNR, Palermo, Italy}
        I.S. Glass; \address{South African Astron. Observ.}
        P. Grandi; \address{ISA/CNR, Frascati, Roma, Italy}
        E. Massaro; \address{Univ. of Rome La Sapienza, Italy} 
        S. Molendi; \address{IFCTR-CNR Milano, Italy}
        P. Padovani; \address{ESA--STScI, Baltimore, USA}
        G.G.C. Palumbo; \address{Univ. of Bologna, Italy}
        C. Perola; \address{Univ. of Rome III, Italy}
        M. Salvati; \address{Arcetri Astroph. Obs., Italy}
        A. Treves; \address{Univ. of Como, Italy}
}

\begin{document}

\begin{abstract}
We report on the BeppoSAX observations of the $\gamma$--ray blazar
PKS 0528+134 performed in Feb and Mar 1997, during a multiwavelength 
campaign involving EGRET and ground based telescopes.
The source was in a faint and hard state, with energy spectral
index $\alpha=0.48 \pm 0.04$ between 0.1 and 10 keV,
and [2--10] keV flux of $2.7\times 10^{-12}$ erg cm$^{-2}$ s$^{-1}$.
No significant variability was observed.
The source was detected in the 20--120 keV band by the PDS, with a flux
lying slightly above the extrapolation from lower X--ray energies.
Comparing this low state with previous higher states of the source,
there is an indication that the X--ray spectrum hardens and the
$\gamma$--ray spectrum steepens when the source is fainter.

\end{abstract}

\maketitle

\section{PKS 0528+134}
PKS 0528+134 ($z=2.07$) is the second most distant quasar detected
by EGRET in the $\gamma$--ray band.
Located in the galactic anticenter region, it is heavily absorbed,
although the estimates of $A_V$ are very uncertain (ranging from 2.3 to 5).
Thus PKS 0528+134 is faint in the optical, with a typical
average magnitude of 19.5 in the $V$ band.
It is a strong and flat radio source, with detected superluminal motion
(with $\beta_{app} \sim 4$, [2]).
Close to Geminga and the Crab, it was frequently observed by EGRET and
seen to flare in March 1993, when the $\gamma$--ray flux was 3 times
brighter than average.
Flux changes in the $\gamma$--band are accompanied by spectral variations
in the sense that the spectrum is harder when brighter [3].

\section{RESULTS}
BeppoSAX observed the source 8 times between Feb 21 and Mar 11, 1997,
within a multiwavelength observing campaign involving also EGRET and
ground based telescopes.

\begin{figure}
\vskip -0.5 true cm
\psfig{file=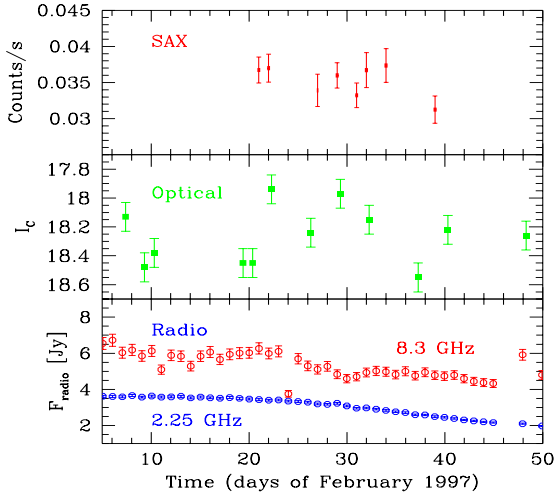,width=8truecm,height=7truecm}
\vskip -1 true cm
\caption[h]{Top panel: BeppoSAX MECS light curve of PKS 0528+134
corresponding to the 8 observations.
Middle panel: the optical light curve in the same period,
in the $I_C$ filter.
Bottom panel: the light curve at 8.3 and 2.25 GHz, by the
Green Bank monitoring campaign.
}
\vskip -0.5 true cm
\end{figure}

\noindent
{\bf Light curves ---}
In Fig. 1 we show the MECS light curve corresponding to the 8 SAX
observations.
The flux is constant (probability of
25\% to obtain a larger value of $\chi^2$, when fitting a constant).
No significant variations are present within the single observations.
On the contrary, small amplitude variations are detected
in the optical and at 8.3 GHz,
as shown in the middle and lower panel of Fig. 1.

\begin{figure}
\vskip -2 true cm
\psfig{file=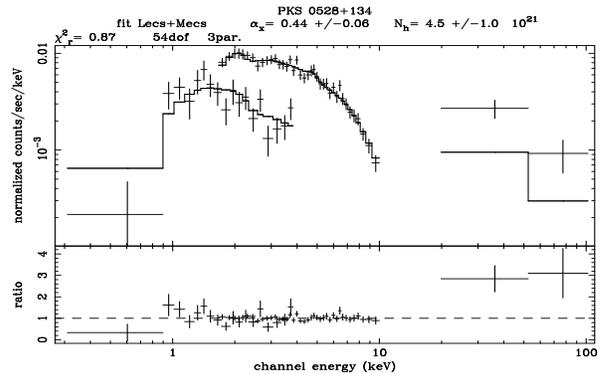,width=8truecm,height=9.truecm}
\vskip -2.5 true cm
\caption[h]{Fit to the LECS and MECS BeppoSAX spectrum of PKS 0528+134.
PDS data are not used for the fit, and appear to lie above the extrapolation
of the power law from lower energies. }
\vskip -1 true cm
\end{figure}

\noindent
{\bf Spectral fitting ---}
Since the flux is constant we decided 
to add together the 8 spectra.
A single power law fit to the LECS+MECS data (see Fig. 2)
yields an energy spectral index $\alpha=0.44\pm 0.06$ and 
$N_H=(4.5\pm 1.0)\times 10^{21}$ cm$^{-2}$, with $\chi^2_r=0.87$ and
54 degree of freedom.
Note that, as shown in Fig. 2, the two PDS data points lye above the
model fit.

A long standing problem with this source is the determination
of its optical extinction and the column $N_H$ along the line of sight.
The value that we found is in good agreement with the sum
of the estimated absorption caused by the column of neutral galactic hydrogen
[$N_H=(2.6\pm 0.1)\times 10^{21}$ cm$^{-2}$],
plus that due to the outer edge of
the molecular cloud Barnard 30 in the $\lambda$ Orion rings of clouds
[$N_H\sim 1.3\times 10^{21}$ cm$^{-2}$].
The best fit value found by published ASCA observations,
$N_H\sim 5\times 10^{21}$ cm$^{-2}$  [3],
is in good agreement with our value.
In addition to the ASCA published spectra, we have re-analyzed
two other ASCA observations performed in 1995, taken from
the ASCA archive.
Fitting the sum of 3 ASCA observations performed in a time interval of
2 weeks, when the source was bright, we found a value of
$N_H=(5.3\pm 0.1)\times 10^{21}$ cm$^{-2}$, which is the best
measure of the column so far.

Since the BeppoSAX measurement is consistent with this value, we decided,
when investigating the flux--slope correlation,
to fix the $N_H$ at this value.
In this case the LECS+MECS data are fitted by a single
power law with an energy index $\alpha=0.48\pm 0.04$.

Including the PDS data and using a single power law fit
we obtain $\alpha=0.44\pm 0.04$, with the PDS points still lying
somewhat above the model fit.
The same data can be fitted with a broken power law model
with $\alpha_1=0.46\pm 0.04$ and $\alpha_2=-0.15\pm 0.29$ and
break energy $E_B\sim 10^{+12}_{-3}$ keV.
The latter model significantly improves the fit, according to the
$F$--test (99\%).

\begin{figure}
\vskip -1 true cm
\psfig{file=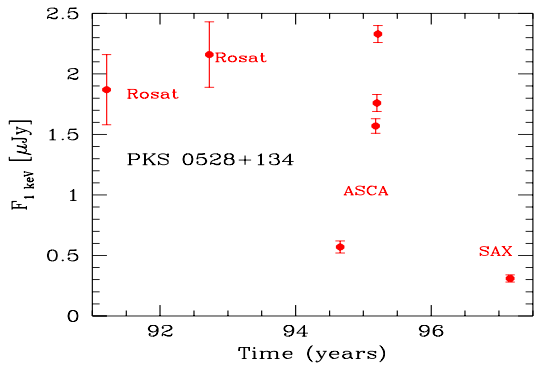,width=8truecm,height=6truecm}
\vskip -1.5 true cm
\caption[h]{Historical light curve in the X--ray band, after 1991.
All data have been reanalyzed assuming a fixed $N_H=5.3\times 10^{21}$
cm$^{-2}$.
}
\vskip -1 true cm
\end{figure}

\noindent
{\bf Historical X--ray light curve ---}
In Fig. 3 we show the light curves of all the observations in the
X--ray band after 1991.
In 1995 the source varied by 50\% in 2 weeks; while
between the 1995 ASCA and our 1997 BeppoSAX observations, when the source was
at its faintest historical level, the flux decreased by a factor $\sim$7.
In the $\gamma$--ray band the source was observed more often, with
a variability of a factor 13 in 2 months, and a factor $\sim$2 in 2 days
during the flare occured in 1993 (see [1]).

\begin{figure}
\vskip -0.7 true cm
\psfig{file=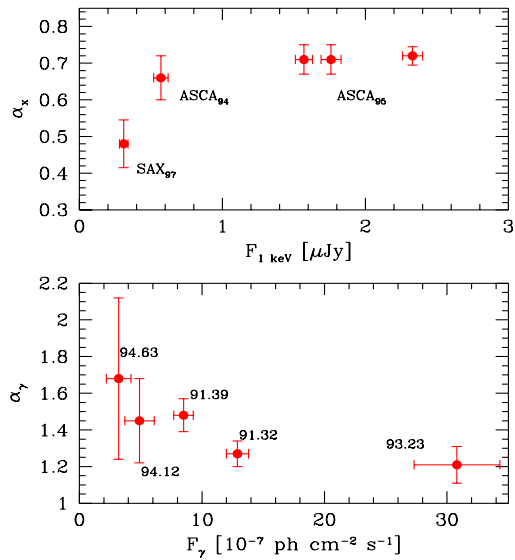,width=8.5truecm,height=8truecm}
\vskip -1.5 true cm
\caption[h]{Spectral indices vs flux in the X--ray and $\gamma$--ray bands.
All X--ray data have been reanalyzed assuming a fixed $N_H=5.3\times 10^{21}$
cm$^{-2}$.
}
\vskip -0.5 true cm
\end{figure}

\noindent
{\bf Flux--spectral index correlations ---}
In Fig. 4 (upper panel) we report the X--ray spectral index
determined by BeppoSAX and ASCA vs the 1 keV flux.
There is an indication of a flattening of the slope when the source
is fainter, just the opposite of what happens in the $\gamma$--ray band
(lower panel).
While in the $\gamma$--ray band this behaviour is similar to
what observed in other sources (see e.g. [4]), the
`flatter when fainter' behaviour in the X--ray band is unusual,
and likely to yield important informations and/or constraints
on the emission models.

\section{THE SED}
In Fig. 5 we show the overall spectral distribution of PKS 0528+134
corrisponding to different epochs, as indicated by the labels.
Data have been dereddened assuming $N_H=5.3 \times 10^{21}$ cm$^{-2}$,
corresponding to $A_V=3$.
As all other $\gamma$--ray bright blazars, also the SED of PKS 0528+134
is characterized by two peaks, one in the far IR region, and the
other one at $\sim$10 MeV.
As discussed in [3], the dereddened optical spectrum
is inverted ($\alpha_o=-0.18\pm 0.08$), and thus indicative of the presence
of the blue-bump.
Note that the MeV--GeV emission dominates the bolometric output by a large
amount, reaching a luminosity in excess of 10$^{49}$ erg s$^{-1}$.

\begin{figure}
\vskip -1 true cm
\psfig{file=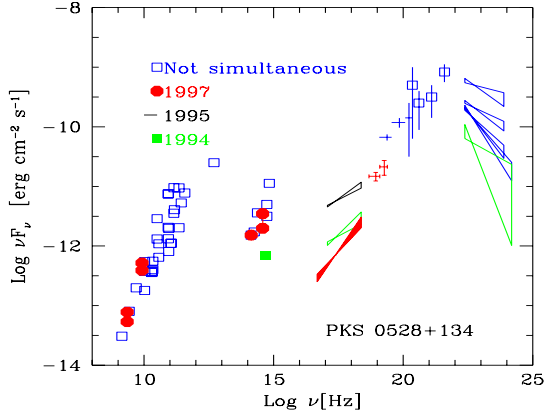,width=8truecm,height=6.5truecm}
\vskip -1.5 true cm
\caption[h]{Overall spectral energy distribution of PKS 0528+134.
Simultaneous observations are marked by the different symbols, as labelled.
Other non--simultaneous data are taken from the literature.
}
\vskip -0.5 true cm
\end{figure}
\section{DISCUSSION}
\noindent
Among the many proposed models for $\gamma$--ray bright blazars, the
most popular are the synchrotron--self Compton (SSC)
and the external Compton (EC) models.
In both of them the synchrotron radiation is responsible for the
low energy emission (up to the optical-UV band), while the inverse
Compton process forms the X-- to the $\gamma$-ray spectrum.
What differs in the two models is the nature of the seed photons
to be Comptonized at high energies.
They are locally produced synchrotron photons in the SSC model, while
in the EC model the main contribution to the seed photons
comes from regions external to the $\gamma$--ray producing zone.
The presence of the blue-bump and of luminous emission lines supports,
in the case of PKS 0528+134, the EC model.
For a detailed discussion of the
fitting of the 1994 overall spectrum with SSC and EC models
see [3].
If confirmed, the variability behaviour in the high energy band
(X--ray spectrum steeper when brighter; $\gamma$--ray spectrum
steeper when fainter) could help to understand the role of the
SSC vs the EC model, and possibly to shed light to the importance
of electron-positron pair production.
We can envisage two scenarios to explain the observed behaviour:

1) The X--rays are produced by $both$ the self Compton and EC processes.
There are $two$ typical frequencies of the seed photons:
one corresponding to the peak of the synchrotron emission
(in the far IR), and one corresponding to the external radiation.
If emission lines and blue bump photons form the bulk of the external
radiation, their typical frequency, as observed in the comoving
frame, is in the far UV.
If the radiation energy densities of these two components
are comparable (within a factor 10), then the self Compton spectrum
will dominate at lower X--ray energies, while the EC spectrum
dominates above, being entirely responsible for the emission
at $\gamma$--ray energies.
This is because the self Compton spectrum is somewhat steeper than the EC 
one produced by the same electrons, and because 
the maximum self Compton frequency is lower.
If the number of emitting electrons increases, then the synchrotron
and the EC flux vary linearly, while the self Compton flux varies
quadratically, making it to dominate the flux over a larger
X--ray energy range.
This could explain the `steeper when brighter' behaviour
of PKS 0528+134 in the X--rays.
To explain the opposite behaviour in the $\gamma$--rays it is necessary
to assume, in addition, that the variation of the electron number is 
accompanied by a flattening of the high energy tail of their energy 
distribution, a behaviour often invoked to explain blazar variability.

2) Another process capable to account for the X--ray
`steeper when brighter' behaviour is e$^\pm$ pair production.
Suppose that, when the source is in a high (and hard) $\gamma$--ray
state, a small fraction of high energy power gets converted in
electron positron pairs, making the particle energy distribution to 
steepen toward lower energies.
This could account for a brighter and steeper medium energy
X--ray spectrum, while at higher energies the emission
of the `normal' particles (with a flatter energy distribution) dominates.
The main difficulty of this scenario is a `fine tuning' problem, since
the high energy power that is absorbed and reprocessed must be of
the right amount (some per cent at most).
If it is less, then pairs are not produced in a sufficient amount
to contribute anywhere in the spectrum.
If it is higher, then they could increase the amount of target photons
for $\gamma$--$\gamma$ collisions, inducing a catastrofic cascade, and an
overproduction of X--ray with respect to what we observe.
A detailed investigation of these ideas is in progress and will be
presented elsewhere.


\end{document}